\documentclass{
JHEP3}
\usepackage{epsfig}

\newcommand{\be}{\begin{equation}}
\newcommand{\ee}{\end{equation}}
\newcommand{\bea}{\begin{eqnarray}}
\newcommand{\eea}{\end{eqnarray}}
\newcommand{\nn}{\nonumber\\ }
\newcommand{\0}{\over }
\newcommand{\2}{{1\over2}}
\newcommand{\4}{{1\over4}}

\def\0{\over } \def\2{{1\over2}} \def\4{{1\over4}}
\def\5{\hat } \def\6{\partial }

\def\g{g_{\rm eff}}

\title{Thermodynamics of Large-$N_f$ QCD at\\
Finite Chemical Potential}

\author{Andreas Ipp and Anton Rebhan\\
Institut f\"ur Theoretische Physik, Technische Universit\"at Wien,\\
  Wiedner Hauptstr. 8--10, A-1040 Vienna, Austria}

\abstract{%
We extend the previously obtained results for the thermodynamic
potential of hot QCD in the limit of large number of fermions to non-vanishing
chemical potential. We give exact results for the thermal pressure
in the entire range of temperature and chemical potential for which
the presence of a Landau pole is negligible numerically. In addition
we compute linear and non-linear quark susceptibilities at zero chemical
potential, and the entropy at small temperatures. 
We compare with the available perturbative results
and determine their range of applicability. Our numerical accuracy
is sufficiently high to check and verify existing results,
including the recent perturbative results by Vuorinen on quark number
susceptibilities and the older results by Freedman and McLerran on
the pressure at zero temperature and high chemical potential. 
We also obtain a number of perturbative coefficients at 
sixth order in the coupling that have not yet been calculated analytically.
In the case of both non-zero temperature and non-zero chemical potential,
we investigate the range of validity of a scaling behaviour
noticed recently in lattice calculations by Fodor, Katz, and Szabo
at moderately large chemical potential and find that it breaks down 
rather abruptly at $\mu_q \gtrsim \pi T$, which points to a presumably generic
obstruction for extrapolating data from small to
large chemical potential.
At sufficiently small temperatures $T\ll \mu_q$, we find dominating
non-Fermi-liquid contributions to the interaction part of the entropy,
which exhibits strong nonlinearity in the temperature and an excess
over the free-theory value.
}

\preprint{TUW-03-13}


\begin{document}

\section{Introduction}

At large temperature and/or chemical potential, one would
expect that asymptotic
freedom should make the deconfined phase of QCD accessible
by analytical methods such as perturbation theory \cite{Kap:FTFT}.
As is well known, the nonperturbative magnetostatic sector
of QCD at high temperature produces a barrier to perturbation
theory, which in the case of the thermal pressure occurs
at order $g^6 T^4$, where $g$ is the strong coupling constant
\cite{Linde:1980ts,Gross:1981br}. However, already well
below the order where this problem occurs, perturbation theory
requires resummations of collective phenomena such as screening.
While these resummations have a well-defined expansion parameter,
the resulting perturbation series exhibit surprisingly poor
convergence behaviour so that certain further resummations
seem to be necessary to make perturbative results useful
at interesting temperatures.
This problem is in fact not specific to QCD, nor to gauge theories
in general, but also occurs in the simplest models such as
scalar field theory. The case of O($N$) $\phi^4$ theory,
which has a solvable  (in fact extremely simple) 
$N\to\infty$ limit has been
used in Ref.~\cite{Drummond:1997cw} to analyse
the properties of thermal perturbation series and as a toy
model for recent proposals of particular partial resummation
methods such as HTL screened perturbation theory 
\cite{Andersen:1999fw,
Baier:1999db,%
Andersen:2002ey
} and
approximations based on 2PI $\Phi$-derivable schemes
\cite{Blaizot:1999ip%
,Blaizot:2000fc,Blaizot:2001vr,Blaizot:2003tw}.
In Ref.~\cite{Moore:2002md}\footnote{For corrected numerical
results see Ref.~\cite{Ipp:2003zr} or the newest hep-ph version
of Ref.~\cite{Moore:2002md}.}, 
the large flavour-number ($N_f$) limit of
QCD has been proposed as a more interesting testing ground
for various methods to overcome the difficulties with
thermal perturbation theory, because unlike O($N\to\infty$) $\phi^4$ theory
large-$N_f$ QCD, though essentially Abelian in its remaining interaction
exhibits many relevant phenomena such as momentum-dependent screening
and damping.\footnote{A similarly complicated but purely scalar
field theory in 6 dimensions, which can also be solved in something similar
to the large-$N_f$ has been studied in Ref.~\cite{Bodeker:1998an},
but being a scalar theory with cubic interactions it involves
instabilities which render a comparison with QCD impossible.}

An exactly solvable theory which involves fermions is however
also of interest with regard to the possibility of exploring
the effects of finite chemical potential, both with respect
to the inherent problems of thermal perturbation theory, and
also beyond. The main nonperturbative method to investigate
real QCD is certainly lattice gauge theory, where recently
important progress has been made to also cover finite chemical
potential \cite{Fodor:2001au,
Fodor:2002km,deForcrand:2002ci,Allton:2002zi,D'Elia:2002gd,Gavai:2003mf},
but the extrapolation to larger chemical potential and smaller
temperature remains uncertain.

In this work we present the extension of the results of
Ref.~\cite{Moore:2002md,Ipp:2003zr} for QCD in the limit of
large $N_f$ to cover the entire range of temperature and
chemical potential for which the problem that large-$N_f$
theory only exists with a cut-off below the scale of the
Landau pole remains negligible numerically. This
is indeed the case when temperature and chemical potential
are sufficiently below the required cut-off, which can be
made exponentially large at small effective coupling.\footnote{The same
issue arises and has been discussed before in the
exactly solvable large-$N$ limits of the scalar models
of Ref.~\cite{Drummond:1997cw,Bodeker:1998an}.}

In particular, we obtain the exact large-$N_f$ result
for the thermal pressure and a number of derived quantities
such as quark number susceptibilities at zero chemical
potential 
and the entropy at small temperatures.
We use these results to compare with known results from
thermal perturbation theory \cite{Arnold:1995eb,Zhai:1995ac,Hart:2000ha}
obtained
at small chemical potential
where dimensional reduction
\cite{Ginsparg:1980ef
,Braaten:1995na,Kajantie:1996dw}
is applicable. Our numerical accuracy turns out
to be sufficiently high to permit the verification of e.g.\
a recent three-loop result of Vuorinen \cite{Vuorinen:2002ue}
on quark number susceptibilities
as well as a numerical coefficient in the pressure
at zero temperature obtained long ago by Freedman and McLerran
\cite{Freedman:1977dm,
Baluni:1978ms}.
We are moreover able to extract a number of perturbative
coefficients at order 
$g^6$ that are not yet known
from analytical calculations.
The comparison of the perturbative results with the exact ones
allows us to investigate the convergence properties and
ambiguities of thermal perturbation theory in some detail.

For the case of both non-zero temperature and non-zero chemical
potential, we are able to explore the range of validity of a scaling behaviour
noticed recently in lattice calculations by Fodor, Katz, and Szabo
\cite{Fodor:2002km}
at moderately large chemical potential and find that it breaks down 
rather abruptly at $\mu_q \gtrsim \pi T$, which points to a presumably generic
obstruction for extrapolating data on the equation of state
from small to high chemical potential.

At small temperatures $T\ll \mu$, we find a region which is dominated
by non-Fermi-liquid behaviour. There the usual linearity of the
entropy in $T$ is replaced by a nonmonotonic behaviour, which
leads to an excess of the entropy over its free-theory value
up to a certain value $T/\mu$ which depends on the strength
of the coupling. Thus, while large-$N_f$ QCD does not have
much in common with the rich phase structure of real QCD, it
allows one to study aspects of non-Fermi-liquid behaviour
\cite{Holstein:1973,Gan:1993,Chakravarty:1995},
which have recently turned out to be of relevance to
the colour superconducting phases of QCD 
\cite{Son:1998uk,Brown:2000eh,Boyanovsky:2000bc,Boyanovsky:2000zj}.

\section{Recapitulation of large-$N_f$ QCD}

In Ref.~\cite{Moore:2002md,Ipp:2003zr} the thermal
pressure of hot
QCD with a large number of fermions $ N_{f}\gg N_{c}\sim 1 $
was calculated exactly
at next-to-leading order (NLO) in a large $ N_{f} $ expansion.
In this limit, the gauge coupling $g^2$ is taken to zero such
that a finite value of $g^2N_f$ of order 1 is obtained. For
the two cases of massless QCD and ultrarelativistic QED the
theory can then be solved exactly, i.e. to all orders in the
effective coupling which following Ref.~\cite{Moore:2002md,Ipp:2003zr}
we define as
\begin{equation}
\g^2 = \left\{
        \begin{array}{cc} 
        \displaystyle \frac{g^2 N_f}{2} \, , & {\rm QCD} \, , \\ & \\
        g^2 N_f \, , & {\rm QED} \, . \\ \end{array} \right.
\end{equation}

At leading
order in $ N_{f} $ the renormalization scale dependence is determined
exactly by the one-loop beta function according to
\begin{equation}
\label{gscal}
\frac{1}{g_{\rm {eff}}^{2}(\mu )}=\frac{1}{g_{\rm {eff}}^{2}(\mu ')}
+\frac{\ln (\mu '/\mu )}{6\pi ^{2}}\,.
\end{equation}

This implies a Landau singularity. Following Ref.~\cite{Moore:2002md}
we define the Landau scale $\Lambda_{\rm L}$ such that the vacuum 
gauge field propagator diverges at $ Q^{2}=\Lambda _{\rm L}^{2} $,
which leads to
\be\label{LLandau}
\Lambda _{\rm L}=\bar\mu_{\rm MS}
 e^{5/6}e^{6\pi ^{2}/\g^{2}(\bar\mu_{\rm MS})}\,.
\ee
The presence of a Landau singularity means that there is
an irreducible ambiguity associated with the UV completion
of the theory, but in the thermal pressure this ambiguity
is suppressed by a factor $({\rm max(T,\mu)}/\Lambda_{\rm L})^4$.

The thermal pressure in the large-$N_f$ limit down to order $N_f^0$ is
diagrammatically given by an undressed fermion loop, which is
of order $N_f^1$, and by a gauge boson loop with an arbitrary number
of fermion loop insertions (plus corresponding counterterm insertions),
which are all of order $N_f^0$ because $g^2N_f \sim O(1)$.
Standard Schwinger-Dyson resummation in the gauge boson loop
includes all diagrams there are to next-to-leading order, $N_f^0$, which
is the order we shall be interested in.

In the imaginary-time formalism the thermodynamic pressure is obtained 
by performing a sum over Matsubara frequencies, which can be replaced
by a contour integral, and subtracting 
the vacuum contribution. The result for the NLO ($N_f^0$) pressure
is \cite{Moore:2002md}
\begin{eqnarray}
\label{PNLO}
\frac{P_{\rm {NLO}}}{N_{g}} & = & 
\int \frac{d^{3}q}{(2\pi )^{3}}\int _{0}^{\infty }\frac{d\omega }{\pi }
\biggr[ \,2\,\Bigl\{ \left[n_{b}+{\textstyle\frac{1}{2}}\right]\,
\rm {Im}\ln \left(q^{2}-\omega ^{2}+\Pi _{T}+\Pi _{\rm {vac}}\right)
\nonumber\\
 &  & \qquad \qquad \qquad \qquad \quad 
-{\textstyle\frac{1}{2}}\,\rm {Im}
\ln \left(q^{2}-\omega ^{2}+\Pi _{\rm {vac}}\right)\Bigr\} \nonumber\\
 &  &  +\,\Bigl\{ \left[n_{b}+{\textstyle\frac{1}{2}}\right]\,
\rm {Im}\ln \frac{q^{2}-\omega ^{2}+\Pi _{L}+
\Pi _{\rm {vac}}}{q^{2}-\omega ^{2}}-{\textstyle\frac{1}{2}}\,
\rm {Im}\ln \frac{q^{2}-\omega ^{2}
+\Pi _{\rm {vac}}}{q^{2}-\omega ^{2}}\Bigr\} \biggr] ,
\end{eqnarray}
where $n_{b}(\omega)=1/(e^{\omega/T}-1)$,
$\Pi _{\rm {vac}}$ is the vacuum part of the gauge-boson
self energy,
\be
\Pi _{\rm {vac}}^{\mu \nu }(Q)=
-\frac{g_{\rm {eff}}^{2}}{12\pi ^{2}}
\left( \eta ^{\mu \nu }Q^{2}-Q^{\mu }Q^{\nu }\right) 
\left( \ln \frac{Q^{2}}{\bar{\mu }^{2}}-\frac{5}{3}\right),
\ee
and $\Pi_T$ and $\Pi_L$ are the two independent structure functions
in the thermal self energy
as given explicitly in Ref.~\cite{Moore:2002md,Ipp:2003zr}.
These cannot be given in closed form (except for their imaginary
parts \cite{Ipp:2003zr}), but can be represented by
one-dimensional integrals involving the fermionic distribution function
$n_f$. 

All that is needed 
to generalize to non-vanishing chemical potential $\mu$ is
to include the latter in the fermionic distribution function appearing
within
the self-energy expressions $\Pi_T$ and $\Pi_L$ according to
\be
n_{f}(k,T,\mu )=\frac{1}{2}\left( \frac{1}{e^{(k-\mu )/T}+1}
+\frac{1}{e^{(k+\mu )/T}+1}\right) .
\ee

When evaluating the integrals above
exactly by numerical means, we can safely integrate parts
proportional to $ n_{b} $ in Minkowski space, since those are
exponentially ultraviolet safe. For terms without $ n_{b} $ 
more care is required, because
the
expressions are potentially logarithmically divergent, unless
a Euclidean invariant cutoff is introduced \cite{Moore:2002md}. As in 
Ref.~\cite{Moore:2002md,Ipp:2003zr} we apply a cutoff and 
stop the $ d^{4}Q $ integration
at $ Q^{2}=a\Lambda _{\rm L}^{2} $, varying the value of
$ a $ between 1/4 and 1/2 to estimate the irreducible ambiguity.

\section{Results and discussion}

\EPSFIGURE[t]{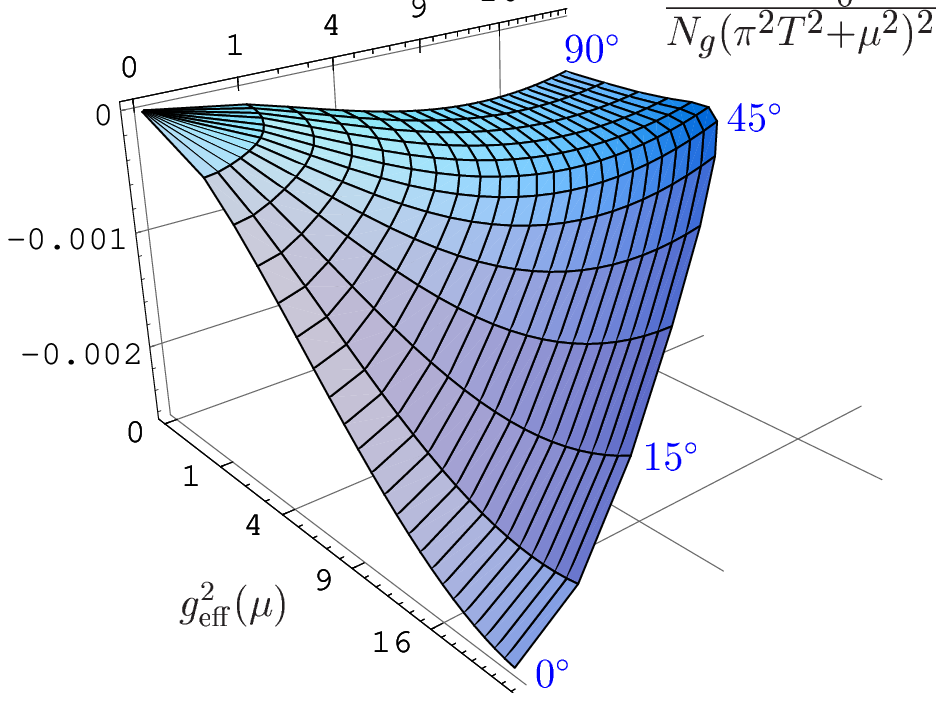
}{\label{fig:3d}
Exact result for the large-$N_f$ interaction pressure $P-P_0$ normalized
to $N_g(\pi^2T^2+\mu^2)^2$ as a function of $\g^2(\bar\mu_{\rm MS})$
with $\bar\mu_{\rm MS}^2=\pi^2T^2+\mu^2$, which is the radial
coordinate, and $\phi=\arctan{\pi T\0\mu}$.
}

In Fig.~\ref{fig:3d} we display our exact results\footnote{Tabulated
results will be made available on-line at
\href{http://hep.itp.tuwien.ac.at/~ipp/data/}{\tt 
http://hep.itp.tuwien.ac.at/~{}ipp/data/} .}
for the
interaction pressure $P-P_0 \propto N_f^0$, where the
ideal-gas limit
\be
P_0=NN_f\left( {7\pi^2 T^4\0180}+{\mu^2 T^2\06}+{\mu^4\012\pi^2} \right)
+N_g{\pi^2T^4\045}
\ee
has been subtracted, for the entire $\mu$-$T$ plane
(but reasonably below the scale Landau pole).
For this we introduce an angle $\phi=\arctan{\pi T\0\mu}$ and
encode the magnitudes $T/\Lambda_{\rm L}$ and $\mu/\Lambda_{\rm L}$
through the running coupling $\g^2(\bar\mu_{\rm MS})$
with $\bar\mu_{\rm MS}^2=\pi^2T^2+\mu^2$ according to (\ref{LLandau}).

We found that the ambiguity arising from the presence of a Landau pole
reaches the percent level for $\g^2\gtrapprox 28$,
where $\Lambda_{\rm L}/\sqrt{\pi^2T^2+\mu^2} \lessapprox 19$.
At larger coupling (corresponding to larger $T$ and/or $\mu$),
this ambiguity grows rapidly and will be shown in the
two-dimensional plots below by a (tiny) red area.

In the following we shall compare the exact large-$N_f$ result
with known results from perturbation theory at high temperature
and small chemical potential, where dimensional reduction
is an effective organizing principle, and with results at
zero temperature, where dimensional reduction does not apply.
We also investigate to what extent quark number susceptibilities
at vanishing chemical potential determine the behaviour at
larger chemical potential.


\subsection{Pressure at small chemical potential}

The perturbative result for the thermal pressure
of hot gauge theories with
fermions has been obtained to order $g^5$ at zero chemical
potential in Ref.~\cite{Arnold:1995eb,Zhai:1995ac} and, using effective
dimensionally reduced field theory, in Ref.~\cite{Braaten:1996jr}.

In the large-$N_f$ limit, dimensional reduction in fact gives
a free theory at order $N_f^0$.
Below order $g^7$, its pressure contribution
is simply $P_{\rm DR}={N_g}m_E^3 T/(12\pi^2)$, where $m_E$
is the Debye mass obtained by integrating out hard momentum
modes to the required perturbative order. 
(At and beyond order $g^7$, the effective theory requires
higher-derivative kinetic terms.)

Dimensional reduction continues to work for small chemical
potential $\mu \lesssim T$. The result to order $\g^5$ for QCD in
the large-$N_f$ limit as available in the literature reads
\bea\label{pintdr}
&&\!\!\!\!\!\!{P-P_0\0N_g}\Big|_{T\gg \mu}=
- \left[ {5\09} T^4 + {2\0\pi^2} \mu^2 T^2 +{1\0\pi^4}\mu^4 
\right]{\g^2\032} + {1\012\pi}Tm_E^3\nn
&&+ \biggl[
\left( {20\03}T^4+{24\0\pi^2}T^2\mu^2 \right) \ln{\bar\mu_{\rm MS}\04\pi T}
+\left( {1\03}-{88\05}\ln 2+4\gamma
-{8\03}{\zeta'(-3)\0\zeta(-3)}+{16\03}{\zeta'(-1)\0\zeta(-1)}
\right) T^4\nn
&&\quad -{26+32\ln 2-24\gamma\0\pi^2}T^2\mu^2 
+
{12\mu^4\0\pi^4}\left[\ln{\bar\mu_{\rm MS}\04\pi T}
+ \gamma + C_4 \right]
+\ldots
\biggr]{\g^4\0(48\pi)^2}
+O(\g^6 T^4),\qquad
\eea
where the terms $\propto \g^4$ and involving $\mu$ have recently been
computed by Vuorinen \cite{Vuorinen:2002ue,Vuorinen:2003fs}. 
The contribution to order $g^5$ follows from the NLO correction
to the effective-field-theory parameter $m_E^2$, computed
at finite $\mu$ in Ref.~\cite{Hart:2000ha}
\be\label{mE2}
{m_E^2\0T^2}=\left({1\03}+{\mu^2\0\pi^2T^2}\right) \g^2
\left\{ 1 - {\g^2\06\pi^2} \left[
\ln{\bar\mu_{\rm MS}\0e^{1/2-\gamma}\pi T}
+{1\02} {\mathcal D}({\mu\0\pi T})
\right] \right\}+O(\g^6)
\ee
with the function
\bea
{\mathcal D}(x)&=&
\int_{-\infty}^{\infty} {dp\0p} \left( {1-e^p \0 1+e^p}
-{1\0e^{p+\pi x}+1}+{1\0e^{-p+\pi x}+1} \right) \nn
&=&\aleph({\textstyle\2})-\aleph({\textstyle\2}(1+ix))=
-2\gamma-4\ln 2-
2\,{\rm Re}\,\psi({\textstyle\2}(1+ix)),
\label{calD}
\eea
where $\psi(z)$ is the digamma function
and $\aleph(z)\equiv \psi(z)+\psi(z^*)$
is notation introduced by Ref.~\cite{Vuorinen:2003fs}.\footnote{The closed-form
result for $\mathcal D$ can be obtained from Ref.~\cite{Vuorinen:2003fs}
after identifying
$$\mathcal D(x)=(4\pi)^2\left[
\tilde \mathcal I^0_2(\mu\!=\!-\pi Tx)-\tilde \mathcal I^0_2(\mu\!=\!0)\right]$$
(cf.~Eq.~(5.4) of \cite{Hart:2000ha} and the definition
(A.2) of \cite{Vuorinen:2003fs}) and using the
result (B.71) 
for $\tilde \mathcal I^0_2$ along with the
definitions (2.15), (2.16), and (2.20) of \cite{Vuorinen:2003fs}.
}
For small $x$ this function can be expanded as
\be
{\mathcal D}(x)= 
4\sum_{n=1}^\infty (-1)^n \left(1-{1\02^{2n+1}}\right) \zeta(2n+1)\, x^{2n}\,,
\ee
evidently with a radius of convergence of 1,
which corresponds to $\mu=\pi T$. %

Because the dimensionally reduced theory is free, the only nonanalytic
terms in $\g^2$ stem from the ``plasmon term'' $\propto m_E^3$.
In particular there are no logarithms $\ln(g)$ which in finite-$N_f$
QCD appear at and beyond order $g^4$ and which have recently
been determined even to order $g^6$ in Ref.~\cite{Kajantie:2002wa}. 

\EPSFIGURE{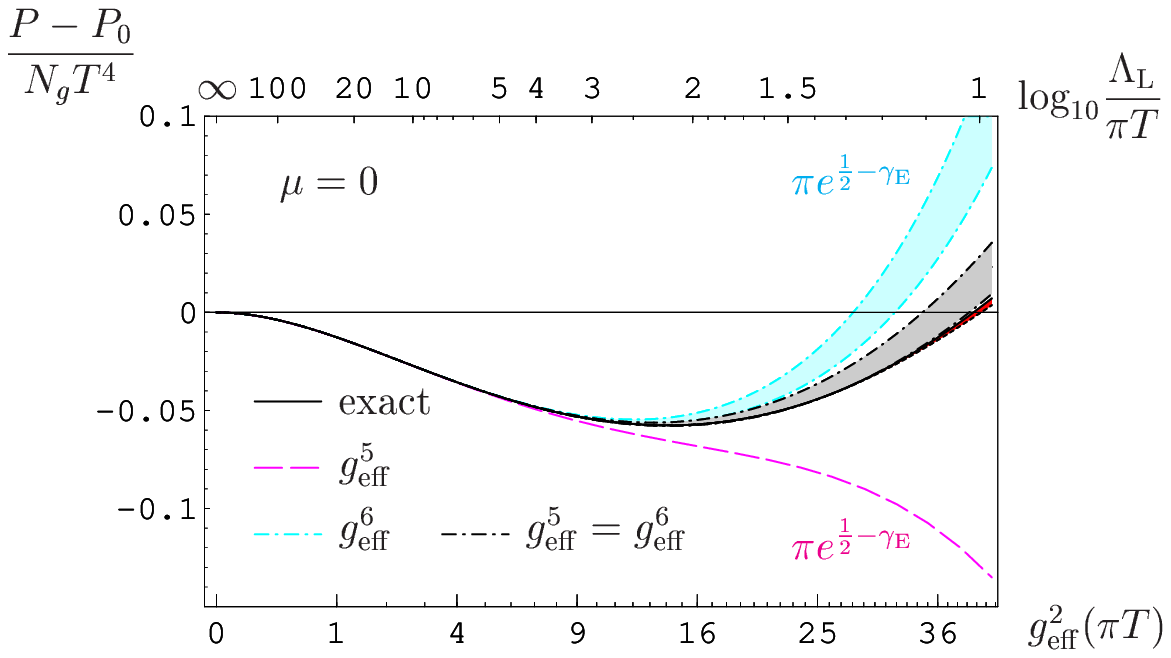
}{\label{fig:zeromu}
Exact result for the interaction pressure at zero
chemical potential as in Ref.~\cite{Moore:2002md,Ipp:2003zr}
but as a function of $\g^2(\bar\mu_{\rm MS}=\pi T)$ or, alternatively,
$\log_{10}(\Lambda_{\rm L}/\pi T)$. The purple dashed line is
the perturbative result when the latter is evaluated
with renormalization scale
$\bar\mu_{\rm MS}=\bar\mu_{\rm FAC}\equiv\pi e^{1/2-\gamma}T$;
the blue dash-dotted lines include the numerically determined
coefficient to order $\g^6$ (with its estimated error)
at the same renormalization scale.
The result marked ``$\g^5=\g^6$'' 
corresponds to choosing $\bar\mu_{\rm MS}$ such
that the order-$\g^6$ coefficient vanishes and retaining
all higher-order terms contained in the plasmon term $\propto m_E^3$.
In this and the following plots the (tiny) red band appearing
around the exact result at large coupling displays the
effect of varying the cut-off from 50\% to 70\% of the
Landau scale $\Lambda_{\rm L}$.}

The numerical result for $\mu=0$, obtained
before in Ref.~\cite{Moore:2002md,Ipp:2003zr},
is shown in Fig.~\ref{fig:zeromu}, but now as a function
of $\g^2(\bar\mu_{\rm MS}=\pi T)$. Also given are the
respective values of $\log_{10}(\Lambda_{\rm L}/\pi T)$.

For small coupling the agreement 
of our numerical results 
with perturbation theory
is sufficiently accurate that it permits a numerical extraction of
the coefficients to order $\g^6$ which are not yet known
analytically. 

Eq.~(\ref{gscal}) dictates that the $\g^6$-term in the pressure
at $\mu=0$ has the form
\bea
{1\0N_g}P\Big|_{\g^6,\mu=0} &=& 
{\left( \frac{\g}{4\pi} \right) }^6 T^4 \left[ C_6
+ 10 \ln^2 \frac{\bar\mu_{\rm MS}}{\pi T}\right. \nn
& &
      -\left. \frac{16 {\pi }^2}{81}
         \left( 1 + 12\gamma  - 
              \frac{464\ln 2}{5} - 
              8\frac{\zeta '(-3)}{\zeta (-3)} + 
              16\frac{\zeta '(-1)}{\zeta (-1)} \right)
              \ln \frac{\bar\mu_{\rm MS} }{\pi T}
         \right]
\eea
and by least-square fitting 
we obtain numerically the estimate
$C_6=+20(2)$.

In real, finite-$N_f$ QCD this result corresponds
to the coefficient involving $N_f^3$ in the
$g^6 T^4$ term of the pressure, which is nonperturbative in its
purely gluonic contribution $\propto N_f^0$.

In Ref.~\cite{Moore:2002md,Ipp:2003zr} the convergence of successive
perturbative approximations to order $\g^5$ has been studied,
with the result that there are large renormalization scale
dependences beyond $\g^2\sim 4$. Fixing this scale dependence
by the requirement of ``fastest apparent convergence'' (FAC)
in the $m_E^2$ parameter of dimensional reduction leads to
$\bar\mu_{\rm MS}=\bar\mu_{\rm FAC}\equiv\pi e^{1/2-\gamma}T$. This choice
leads to fairly accurate results up to $\g^2\sim 9$.

At still larger coupling the exact result for the pressure
at zero chemical potential has the remarkable feature of
bottoming out and tending toward the ideal-gas limit. 
Before it begins to exceed the latter, however, 
the Landau pole starts to influence the result
noticeably. This is
displayed in the figures by the tiny red area around
the ``exact'' result, which
represents the effect of varying the cut-off from 70\% to 50\%
of the Landau scale $\Lambda_{\rm L}$. (See Ref.~\cite{Ipp:2003zr}
for the effect of varying the cut-off independently in
the Minkowski and Euclidean parts of the numerical integrations.)

Including our numerical estimate of the $\g^6$-coefficient
and using $\bar\mu_{\rm FAC}$ further improves the
perturbative result so that it remains accurate up to $\g^2\sim 16$.
The agreement with the exact result can even be further
improved by fixing the renormalization point such that
the $\g^6$ coefficient vanishes and keeping all orders
of the odd terms in $\g$ by leaving the plasmon term $\propto m_E^3$
unexpanded in $\g^2$. The result of this procedure is indicated
by the gray area in Fig.~\ref{fig:zeromu}.
The apparent success is in line with the recent observation
in Ref.~\cite{Blaizot:2003iq} that keeping the parameters
of the dimensionally reduced theory unexpanded greatly improves
the convergence of thermal perturbation theory.

\subsection{Quark number susceptibilities}

\subsubsection{Linear quark number susceptibility}

The (linear) quark number susceptibility is defined as the
first derivate of the quark number density ${\mathcal N}$ with respect
to chemical potential, 
\be
\chi={\6\mathcal N\0\6\mu}={\6^2 P\0 \6 \mu^2}.
\ee

Fig.~\ref{fig:chi} displays the exact large-$N_f$ result for the interaction
part of $\chi$ at zero chemical potential as a function
of $\g$ (or alternatively $\log_{10}(\Lambda_{\rm L}/\pi T)$). Similar
to the thermal pressure, the result is nonmonotonic, but
the minimum already occurs at $\g^2(\pi T)\approx 8.6$, 
and the free-theory value
is recovered at $\g^2(\pi T)\approx 22.5$, where the Landau ambiguity is still
well under control since $\Lambda_{\rm L}/T\approx 100$ at that
coupling.

\EPSFIGURE[t]{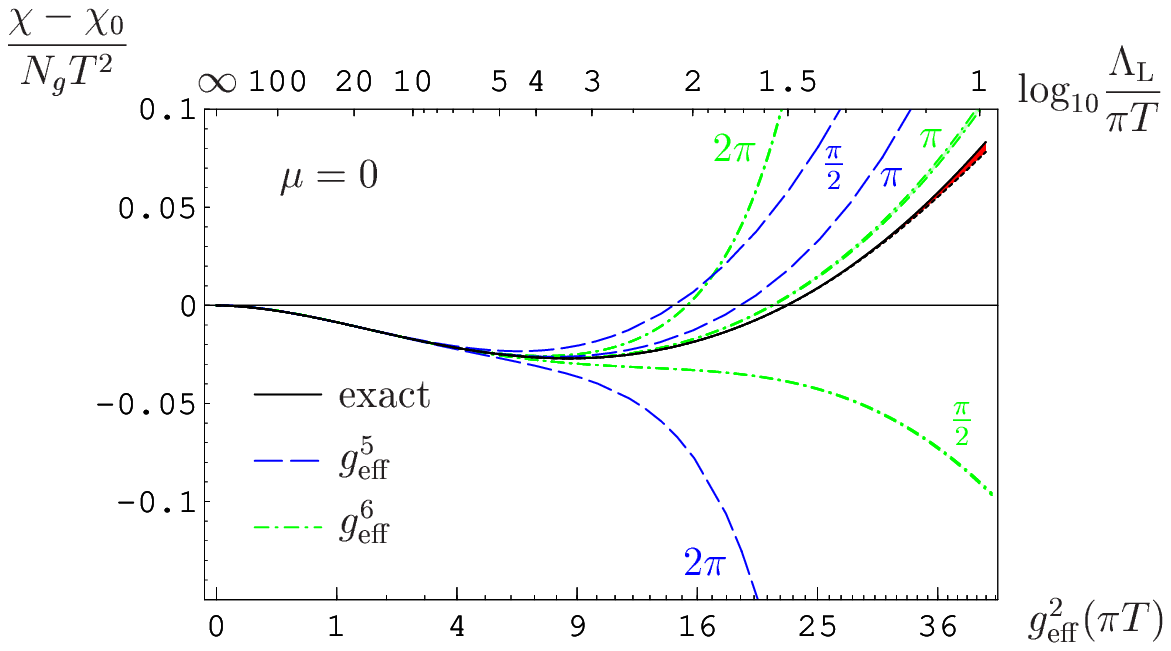
}{\label{fig:chi}
The interaction part of the quark number susceptibility at $\mu=0$
compared with strict perturbation theory to order $\g^5$ and
$\g^6$, respectively, with renormalization scale varied about
$\pi T$ by a factor of 2. 
}

The perturbative (dimensional reduction) result can be read from the
linear term in $\mu^2$ of (\ref{pintdr}) and gives
\bea
 {\chi-\chi_0\0N_gT^2} &=& 2{\6\0\6(\mu^2)}
{P-P_0\0N_g}\Big|_{\mu=0}
=-{\g^2\08\pi^2}+{\g^3\04\pi^3\sqrt3} \nn
&&+{\g^4\048\pi^4}\left[\ln{\bar\mu_{\rm MS}\04\pi e^{-\gamma} T}
-{13\012}-{4\03}\ln2 \right]\nn&&
+{\g^5\016 \pi^5\sqrt3}
\left[-\ln{\bar\mu_{\rm MS}\0e^{1/2-\gamma}\pi T}+{7\018}\,\zeta(3) \right].
\eea

The coefficient of $\g^4$ has only recently been obtained
in \cite{Vuorinen:2002ue} in a three-loop calculation. 
We can confirm its closed-form value
by a numerical fit, which gives agreement with an accuracy of
$2\times 10^{-4}$, thus providing a good check on both our numerics
and the analytical calculations of \cite{Vuorinen:2002ue}. This
level of accuracy allows us to also extract the
order-$g^6$ term as (for $\mu_{\rm MS}=\pi T$)
\be
{\chi|_{\g^6}\0N_gT^2}
 = -4.55(9)\times \left({\g\04\pi}\right)^6.
\ee

In Fig.~\ref{fig:chi} we show the perturbative results to order
$\g^5$ and $\g^6$,
varying the renormalization scale about $\pi T$ by a factor of 2
(now without the improvement of keeping effective-theory parameters
unexpanded). The value $\bar\mu_{\rm MS}=\pi T$ is in fact close
to $\bar\mu_{\rm FAC}$ where it makes no difference whether
$m_E^2$ is kept unexpanded or not. We find that the quality of the perturbative
result for the susceptibility is comparable to that observed in the pressure,
with $\bar\mu_{\rm MS}=\pi T$ being close to the optimal choice.

\subsubsection{Higher-order quark susceptibility}

We have also computed explicitly the 
higher-order susceptibility $\6^4 P/\6 \mu^4|_{\mu=0}$ (which has
recently been investigated in lattice QCD with
$N_f=2$ in Ref.~\cite{Gavai:2003mf}).

\EPSFIGURE[t]{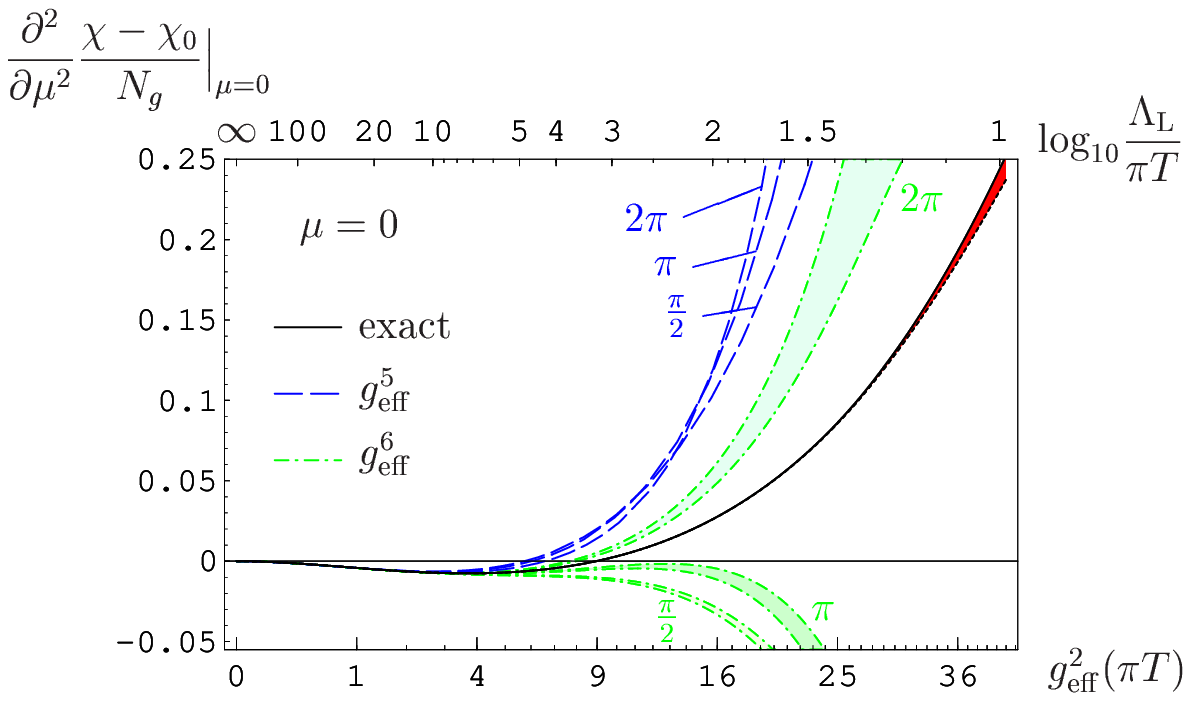
}{\label{fig:chi4}
The interaction part of the higher-order quark number susceptibility 
$\6_\mu^2(\chi-\chi_0)$ at $\mu=0$
compared with strict perturbation theory to order $\g^5$ and
$\g^6$, respectively, with renormalization scale varied about
$\pi T$ by a factor of 2. The coloured bands of the $\g^6$-results
cover the estimated
error of the numerically extracted perturbative coefficients.
}

Our exact result in the large-$N_f$ limit is shown in Fig.~\ref{fig:chi4}.
In this quantity, we find that
the nonmonotonic behaviour observed above in the pressure
and the linear susceptibility is much more pronounced. The minimum now occurs
at $\g^2 \approx 3.7$,
where perturbation theory is still in good shape, and
the free-theory value is exceeded for $\g^2 \gtrapprox 9$.
Using (\ref{pintdr}) we find to order $g^5$
\bea\label{chinonlinpert}
&&{\6^2\0\6\mu^2}{\chi-\chi_0\0N_g}\Big|_{\mu=0}
= {\6^4\0\6\mu^4}{P-P_0\0N_g}\Big|_{\mu=0}
 = 12{\6^2\0(\6\mu^2)^2}{P-P_0\0N_g}\Big|_{\mu=0}
\nn
&=&-{3\g^2\04\pi^4}+{3\sqrt3 \g^3\04\pi^5}
+{\g^4\08\pi^6}\left[\ln{\bar\mu_{\rm MS}\04\pi T}
+ \gamma + C_4 \right]\nn&& +{3\sqrt3 \g^5\016 \pi^7}
\left[-\ln{\bar\mu_{\rm MS}\0e^{1/2-\gamma}\pi T}+{7\03}\,\zeta(3)
-{31\054}\,\zeta(5) \right]+O(\g^6).
\eea

In a first version of this paper
we have extracted the coefficient appearing at order $\g^4$
numerically as $C_4=-7.02(3)$. In the meantime, the complete
$\mu$ dependence of the dimensional reduction result to
order $\g^4$ has been worked out in Ref.~\cite{Vuorinen:2003fs}
from where one can obtain the exact result
\be\label{C4exact}
C_4 = -\frac{1}{12} - 12 \ln2 + \frac{7}{6} \zeta(3) = -6.9986997796998\ldots
\ee
The complete agreement with Ref.~\cite{Vuorinen:2003fs}
provides on the one hand an independent check on the correctness
of the 3-loop calculations of \cite{Vuorinen:2003fs} and
on the other hand a check on the accuracy of our
numerical analysis.

Using (\ref{C4exact}) we can extract the
term of order $\g^6$ in (\ref{chinonlinpert}) as $-39(1)\g^6/(128\pi^8)$
for $\bar\mu=\pi T$. The perturbative results to order $\g^5$
and to order $\g^6$ are compared with the exact result in Fig.~\ref{fig:chi4}.
This shows that
the accuracy of the perturbative result again improves by
going from order $\g^5$ to order $\g^6$, but the renormalization
scale dependence increases sharply at large coupling.\footnote{%
The size of the scale dependence can in fact be reduced somewhat
by keeping odd powers of $m_E$ without expanding out the $\g^4$
correction in $m_E^2$.}

\subsubsection{Pressure at larger chemical potential from susceptibilities}

With regard to the 
recent attempts to explore 
QCD at finite
chemical potential by
means of lattice gauge theory 
\cite{Fodor:2002km,deForcrand:2002ci,Allton:2002zi,D'Elia:2002gd},
it is of interest how well
the pressure at larger chemical potential can be approximated
by 
the first few terms of a Taylor series in $\mu^2$.

In Ref.~\cite{Fodor:2002km} it has been observed that
the ratio of $\Delta P=P(T,\mu)-P(T,\mu=0)$ over the
corresponding free-theory quantity $\Delta P_0$ is practically
independent of $\mu$ for the range of chemical potentials
explored. This is also realized when quasi-particle models
are used for a phenomenological extrapolation of lattice data
\cite{Szabo:2003kg,Rebhan:2003wn} in a method introduced by Peshier et al. 
\cite{Peshier:1999ww
}.

\EPSFIGURE[t]{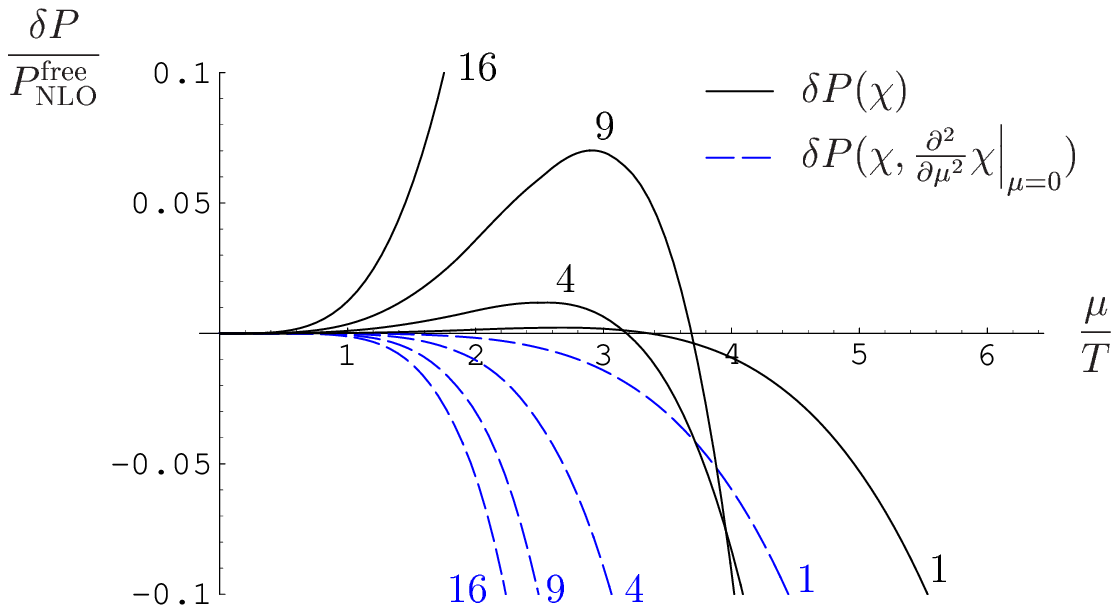}{\label{fig:deltap}
Deviation from the scaling observed in Ref.~\cite{Fodor:2002km}
in lattice QCD for small chemical potential
in the quantity
$\delta P=P(T,\mu)-P(T,0)-{1\02}\chi|_{\mu=0} 
(\mu^2+\mu^4/(2\pi^2
T^2))$ (full lines) and
in $\delta P=P(T,\mu)-P(T,0)
-{\mu^2\02}\chi|_{\mu=0}  - {\mu^4\04!}{\6^4P\0\6\mu^4}|_{\mu=0}$ 
(dashed lines),
both
normalized to $P_{\rm NLO}^{\rm free}=N_g \pi^2 T^4/45$,
for $\g^2(\pi T)=1,4,9,16$.}

In Fig.~\ref{fig:deltap} we show the deviation from this
``scaling'' at higher values of $\mu/T$
by considering the quantity
$\delta P=P(T,\mu)-P(T,0)
-{1\02}\chi|_{\mu=0} (\mu^2+\mu^4/(2\pi^2
T^2))$ divided by $P_{\rm NLO}^{\rm free}=N_g \pi^2 T^4/45$.
The combination $(\mu^2+\mu^4/(2\pi^2 T^2))$ 
appearing therein is such that 
a replacement of $P$ and $\chi$ by their interaction-free values
$P_0$ and $\chi_0$ makes $\delta P$ vanish identically. 
(As can be seen from
the above perturbative results, $\delta P$ also vanishes
for the leading-order interaction parts $\propto \g^2$.)
In the exact large-$N_f$ results of Fig.~\ref{fig:deltap} we observe that
for coupling $\g^2 \lesssim 4$ the deviation $\delta P$ is at most a few
percent of $P_{\rm NLO}^{\rm free}$ for $\mu/T\lesssim \pi$, but
it rapidly grows for $\mu/T \gtrsim \pi$.
This is in fact also nicely illustrated by the 3-dimensional plot
of the pressure in Fig.~\ref{fig:3d}, which has a rather
conspicuous kink at $\phi=45^\circ$ corresponding to $\mu=\pi T$.

It turns out that including the exact higher-order susceptibility
at $\mu=0$ does not lead to a better approximation
of the pressure at larger chemical potential.
The dashed lines in Fig.~\ref{fig:deltap} correspond
to $\delta P=P(T,\mu)-P(T,0)
-{\mu^2\02}\chi|_{\mu=0}  - {\mu^4\04!}{\6^4P\0\6\mu^4}|_{\mu=0}$. 
While this slightly improves matters at 
small $\mu/T$, it results into even quicker deviations
for larger $\mu/T$.

It is of course impossible to say whether this behaviour
would also appear in real QCD, but since it occurs already at
comparatively small $\g$ in the large-$N_f$ limit,
where
the peculiar nonmonotonic behaviour of the pressure as a function
of $\g$ does not yet arise (the minimum in the normalized interaction
pressure occurs at $\g^2(\pi T)\approx 14$),\footnote{%
Peshier \cite{Peshier:2002fm} has recently argued that
the strong-coupling behaviour of large $N_f$ QCD will be
relevant for real QCD at most in the coupling range where
the normalized pressure decreases with $\g$.}
it may be taken as an indication
that extrapolations of lattice data on the equation of state from 
small chemical
potential to large $\mu/T$ 
are generally problematic.
If anything, real QCD should be more complicated because
of the existence of phase transitions which are absent at NLO in the
large-$N_f$ limit.

\subsection{Pressure at zero temperature}

Our exact result for the thermal pressure at zero temperature 
and finite chemical potential is given in Fig.~\ref{fig:zerot}
as a function of $\g^2(\bar\mu_{\rm MS}=\mu)$. In contrast
to the pressure at zero chemical potential and finite temperature,
the interaction pressure divided by $\mu^4$ is monotonically
decreasing essentially all the way up to the point where the Landau ambiguity
becomes noticeable.

\EPSFIGURE{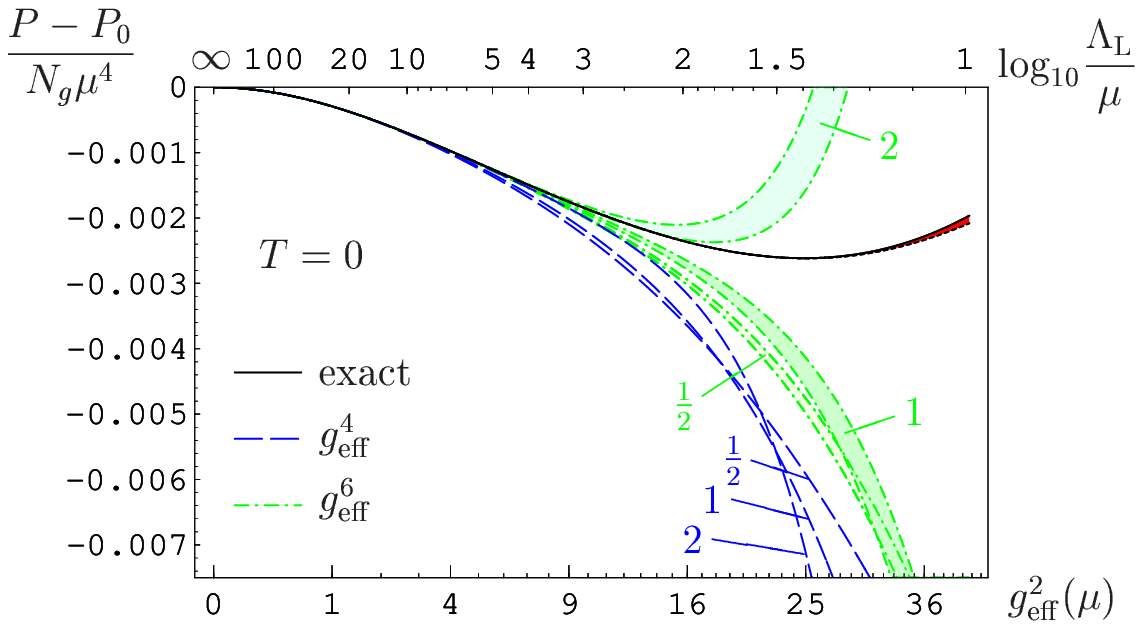
}{\label{fig:zerot}
The interaction part of the pressure at zero temperature and
finite chemical potential as a function of $\g^2(\bar\mu_{\rm MS}=\mu)$
or, alternatively, $\log_{10}(\Lambda_{\rm L}/\mu)$,
compared with the perturbative result of Freedman and McLerran
\cite{Freedman:1977dm,
Baluni:1978ms}
to order $\g^4$, and our numerically extracted order-$\g^6$
result, both with renormalization scale in the perturbative
results varied around $\bar\mu_{\rm MS}=\mu$ by a factor of 2.
The coloured bands of the $\g^6$-results
cover the error of the numerically extracted perturbative coefficients.
}

The thermal pressure at zero temperature and large chemical potential
for QED and QCD has been obtained to order $g^4$ long ago by 
Freedman and McLerran
\cite{Freedman:1977dm,
Baluni:1978ms}.
At this order, there is a non-analytic zero-temperature plasmon
term $\propto g^4 \ln(g)$, whose prefactor is known exactly, but
the constant under the logarithm only numerically.
The transposition of their result, which has been obtained
in a particular momentum-subtraction scheme, to the gauge-independent 
$\overline{\hbox{MS}}$ scheme can be found in 
Refs.~\cite{Blaizot:2000fc,Fraga:2001id}. The large-$N_f$ limit
of this result reads
\be\label{PFMcL}
{P-P_0\0N_g\,\mu^4 }\Big|_{T=0}=
-{\g^2\032\pi^4}-\left[\ln{\g^2\02\pi^2}-{2\03}\ln{\bar\mu_{\rm MS}\0\mu}
-\tilde C_4\right]
{\g^4\0128\pi^6}+O(\g^6\ln\g^{\phantom{6}})
\ee
and involves one of the numerical constants computed in
Ref.~\cite{Freedman:1977dm
},
\be
\tilde C_4=
\frac{79}{18} - \frac{{\pi }^2}{3} - \frac{7\,\log (2)}{3}
- \frac{2\,b}{3} \approx 0.536,
\ee
where $b$ has an integral representation, given in Eq.~(II.3.25)
of Ref.~\cite{Freedman:1977dm}, that apparently cannot
be evaluated in closed form. Fixing an obvious typo\footnote{Comparison
with Eq.~(II.3.24)
shows that there is
a missing exponent 2 after the second set of large round parenthesis
in Eq.~(II.3.25) of \cite{Freedman:1977dm}.}
in Eq.~(II.3.25)
of \cite{Freedman:1977dm}, $b$ can however
be easily evaluated numerically to higher
accuracy than that given in \cite{Freedman:1977dm} as
$b=-1.581231511\ldots$, which leads to $\tilde C_4=0.5358316747\ldots\,$.

The accuracy of our numerical results is sufficiently high
to confirm 
the correctness of the result for $\tilde C_4$
with an accuracy of $\sim 2\times 10^{-4}$.
With the knowledge of the exact value of $\tilde C_4$ we can
also extract, with 
lower precision, the next coefficients
at order $\g^6$, which again involve a logarithmic term:
\bea
{P-P_0\0N_g\,\mu^4 }\Big|_{T=0}&=&
-{\g^2\032\pi^4}-\left[\ln{\g^2\02\pi^2}-{2\03}\ln{\bar\mu_{\rm MS}\0\mu}
-\tilde C_4\right]{\g^4\0128\pi^6}\nonumber\\&& \qquad
- \biggl[\left(
3.18(5)-{16\03}\ln{\bar\mu_{\rm MS}\0\mu}\right)
\ln{\g^2\02\pi^2}
\\&& 
+{16\09}\ln^2{\bar\mu_{\rm MS}\0\mu}
+{16\03}\left(\tilde C_4-{1\02}\right)\ln{\bar\mu_{\rm MS}\0\mu}
-3.4(3)
\biggr]{\g^6\02048\pi^8}
+\ldots\nonumber
\eea

In Fig.~\ref{fig:zerot} we also study the renormalization scale
dependence and apparent convergence
of the perturbative result. 
We have varied $\bar\mu_{\rm MS}$ about $\mu$ by a factor of 2,
and it emerges that the larger values are somewhat favoured.

At low temperature $T\ll\mu$, dimensional reduction does not
occur. If one nevertheless considers the effective-field-theory
parameter $m_E^2$ of (\ref{mE2}) in this limit, one finds
that the function 
${\mathcal D}(x)$ therein approaches
$
-2\,(\ln 2x+\gamma  
)$, so that the $T\to0$ limit of $m_E^2$ exists and reads
\be
m_E^2\to \mu^2{\g^2\0\pi^2} \left\{ 1 - {\g^2\06\pi^2} \left[
\ln{\bar\mu_{\rm MS}\02\mu}-\2 
\right] \right\}+O(\g^6).
\ee
Fastest apparent convergence applied to this quantity would
suggest $\bar\mu_{\rm MS} =2e^{\2}\mu \approx 3.3 \mu$. This turns out
to be not as good as the choice of $2\mu$, though slightly better
than $\bar\mu_{\rm MS}=\mu$.

\subsection{Entropy at small temperatures and non-Fermi-liquid behaviour}

The effect of small temperature on the pressure at
nonzero chemical potential can be studied in terms of the entropy
\be
\mathcal S = \left(
{\6 P\0\6 T} \right)_\mu\,,
\ee
from which
the specific heat can be derived.
(Both, the entropy and the specific heat vanish in the zero-temperature
limit, and the various kinds of specific heat have the same
small-$T$ behaviour up to terms $\sim T^2$.)

At small temperature $T\ll \mu$ one might expect the
contributions involving the Bose-Einstein distribution $n_b$
in the thermodynamic potential (\ref{PNLO}) to be
negligible compared to the ``non-$n_b$'' contributions. 

Considering the latter contributions first, we find that the
corresponding part of the
entropy vanishes linearly
\be\label{Tsigma}
\mathcal S_{\mbox{\scriptsize non-}n_b} \to T \sigma \quad \mbox{for $T\to 0$}
\ee
with
\be
 {\sigma-\sigma_0\0N_g\mu^2} = {\6^2\0\6T^2}
{P_{\mbox{\scriptsize non-}n_b}-P_0\0N_g \mu^2}\Big|_{T=0}
=-{\g^2\08\pi^2}+O(\g^4 \ln \g).
\ee
The coefficient at order $\g^2$ is in accordance with the $\g^2$ part
of the strictly perturbative result for the pressure, which is also known
as the exchange term \cite{Kap:FTFT} and which 
coincides with the $\g^2$ part of (\ref{pintdr}).

At small coupling, we can also extract
the order-$g^4 \ln(g)$ corrections to $\sigma$ from 
$\mathcal S_{\mbox{\scriptsize non-}n_b}$ numerically
with the result
\be
 {1\0N_g\mu^2}\,\sigma\big|_{\g^4}=
{\g^4\032\pi^4}\left[{2\03}\ln{\bar\mu_{\rm MS}\0\mu}
-0.328(1)\times \ln{\g^2\02\pi^2} + 0.462(5) \right]
+\ldots
\ee

The exact result for $\mathcal S_{\mbox{\scriptsize non-}n_b}$ 
is given by the dash-dotted lines in Fig.~\ref{fig:sst}
for $\g^2(\bar\mu_{\rm MS}\!=\!\mu)=1$, $4$,
and $9$ and $0<T/\mu<0.15$.
In this range of temperatures, $\mathcal S_{\mbox{\scriptsize non-}n_b}$
is well approximated by the linear term (\ref{Tsigma}).

\EPSFIGURE{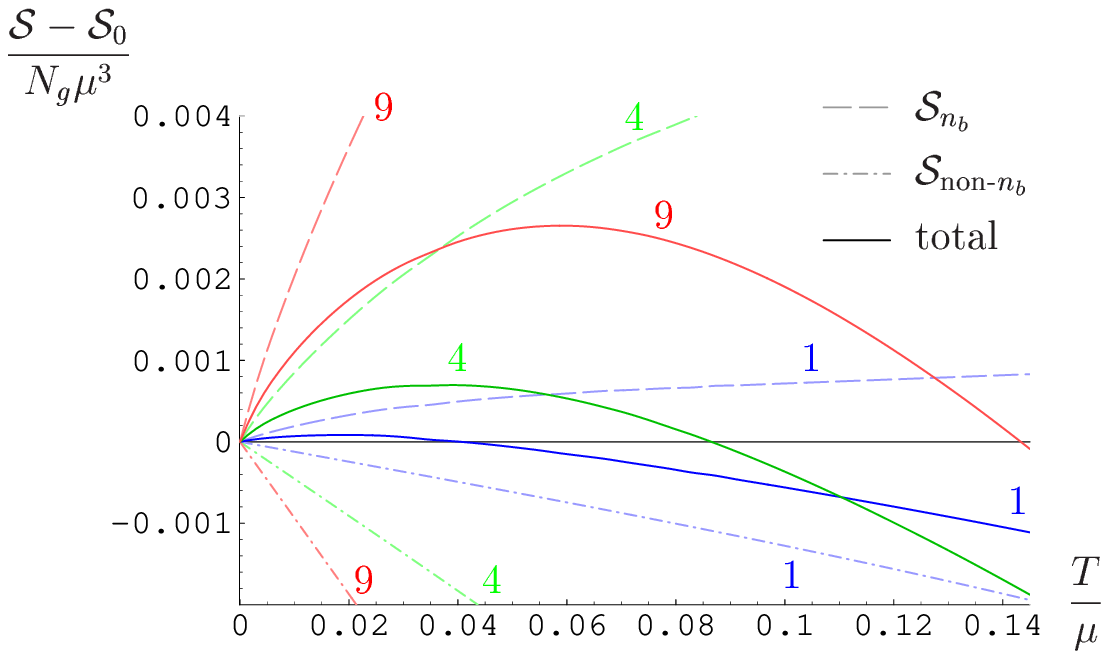%
}{\label{fig:sst}
The interaction part of the
entropy at small $T/\mu$ for $\g^2(\bar\mu_{\rm MS}\!=\!\mu)=1$, $4$,
and $9$.
The ``non-$n_b$'' contributions (dash-dotted lines)
are negative and approximately linear
in $T$ with a coefficient agreeing with the exchange term $\propto \g^2$
in the pressure at small coupling; the ``$n_b$'' contributions 
(dashed lines), which are
dominated by transverse gauge boson modes, are
positive and nonlinear in $T$ such that the total entropy exceeds
the free-theory value at sufficiently small $T/\mu$.
}

Numerically evaluating also $\mathcal S_{n_b}$, i.e.\
the contributions to the entropy following from the parts of
(\ref{PNLO}) which involve $n_b$, we find that these
cannot be neglected at small
temperatures. As shown
in  Fig.~\ref{fig:sst}, for sufficiently small $T/\mu$,
$\mathcal S_{n_b}$ is positive and
even dominates so that the total result
for the entropy turns out to {\em exceed} its free-theory value
for a certain range of $T/\mu$, which gets larger with increasing $\g^2$.

The largest part of the positive and nonlinear
contributions at small $T/\mu$ in fact comes from the transverse
vector-boson modes. At small frequencies, these 
are only weakly dynamically screened (and completely unscreened
in the static limit because of gauge invariance).
As has been discussed in a variety of contexts
(nonrelativistic and relativistic QED as well as colour superconducting QCD) 
in Refs.~\cite{Holstein:1973,Gan:1993,Chakravarty:1995,Son:1998uk,Brown:2000eh,Boyanovsky:2000bc,Boyanovsky:2000zj},
this fact gives rise to non-Fermi-liquid behaviour
at sufficiently small temperature.

A particular consequence is the appearance of anomalous contributions
to the entropy as well as specific heat. 
In Refs.~\cite{Holstein:1973,Gan:1993,Chakravarty:1995}
these contributions have been found to be of the order $-\g^2\mu^2 T\ln T$.
This has however been
questioned recently by the authors of Ref.~\cite{Boyanovsky:2000zj},
who found a different behaviour which in the large-$N_f$ limit
would imply $-\g^2 T^3 \ln T$ as the dominant non-Fermi-liquid contribution,
with $T \ln T$ contributions appearing only at higher order in the coupling.
Because in both of these apparently conflicting results there are
also non-anomalous and not yet determined terms such as $\g^2 \mu^2 T$,
we cannot discriminate with certainty between the two.
On the other hand, 
we could verify that the coefficient of the $-T\ln T$ contribution
as given in \cite{Chakravarty:1995} has
about the correct magnitude to permit a reasonable fit to
our exact results, favouring in fact a leading power of $\g^2$
in this contribution for $T/\mu \lesssim \g^2/(4\pi^2)$ (where the
analysis of Ref.~\cite{Boyanovsky:2000zj} may well cease to be applicable).

We intend to investigate this matter however in more detail in a separate work.

\section{Conclusion}

We have extended the previously obtained exact result for
the pressure of hot QCD in the limit of large flavour number
to finite chemical potential.
For small coupling we have been able to confirm numerically
a number of previously calculated perturbative coefficients 
with great accuracy, and have even obtained numerical values
for a few perturbative coefficients at order 
$g^6$ which were not yet known. At larger coupling we
have studied the applicability
of the perturbative results, their renormalization scale
dependence and apparent convergence at the highest available order in the
coupling.

For the region of simultaneously non-zero temperature and chemical
potential, we have observed a comparatively weak dependence of the
pressure on the chemical potential for $\mu<\pi T$, which is
correspondingly well described by the quark number susceptibility at
zero chemical potential. At $\mu \sim \pi T$, we instead
observed a rather abrupt change leading to a breakdown
of the small-$\mu$ scaling. We suspect that also in
real QCD, where a simple scaling determined by the
quark number susceptibility has been observed recently in lattice
QCD \cite{Fodor:2002km}, 
this may be similarly limited.

At small temperatures $T\ll \mu$ we found non-Fermi-liquid contributions
to the entropy. As a result, for a certain range of temperature,
entropy and specific heat show
strong deviations from linearity in $T$ and also an excess over
their free-theory values.
The exactly solvable large-$N_f$ limit of QED and QCD evidently allows one
to study the phenomenon of non-Fermi-liquid behaviour
beyond perturbation theory. 
A more detailed investigation
will be the subject of a future publication.

\acknowledgments

We would like to thank Dan Boyanovsky and Hector de Vega
for correspondence, and our anonymous referee
for pointing out to us the closed-form result for the
funtion $\mathcal D$ in Eq.~(\ref{calD}). 
This work has been supported by the Austrian Science Foundation FWF,
project no. 14632-TPH.


\begin{thebibliography}{10}

\bibitem{Kap:FTFT}
J.~I. Kapusta, {\em Finite-temperature field theory}.
\newblock Cambridge University Press, Cambridge, UK, 1989.

\bibitem{Linde:1980ts}
A.~D. Linde, {\it Infrared problem in thermodynamics of the {Y}ang-{M}ills
  gas},  {\em Phys. Lett.} {\bf B96} (1980) 289.

\bibitem{Gross:1981br}
D.~J. Gross, R.~D. Pisarski, and L.~G. Yaffe, {\it {QCD} and instantons at
  finite temperature},  {\em Rev. Mod. Phys.} {\bf 53} (1981) 43.

\bibitem{Drummond:1997cw}
I.~T. Drummond, R.~R. Horgan, P.~V. Landshoff, and A.~Rebhan, {\it Foam diagram
  summation at finite temperature},  {\em Nucl. Phys.} {\bf B524} (1998)
  579, [\href{http://xxx.lanl.gov/abs/hep-ph/9708426}{{\tt
  hep-ph/9708426}}].

\bibitem{Andersen:1999fw}
J.~O. Andersen, E.~Braaten, and M.~Strickland, {\it Hard-thermal-loop
  resummation of the free energy of a hot gluon plasma},  {\em Phys. Rev.
  Lett.} {\bf 83} (1999) 2139,
  [\href{http://xxx.lanl.gov/abs/hep-ph/9902327}{{\tt hep-ph/9902327}}];
{\it Hard-thermal-loop
  resummation of the thermodynamics of a hot gluon plasma},  {\em Phys. Rev.}
  {\bf D61} (2000) 014017, [\href{http://xxx.lanl.gov/abs/hep-ph/9905337}{{\tt
  hep-ph/9905337}}];
{\it Hard-thermal-loop
  resummation of the free energy of a hot quark-gluon plasma},  {\em Phys.
  Rev.} {\bf D61} (2000) 074016,
  [\href{http://xxx.lanl.gov/abs/hep-ph/9908323}{{\tt hep-ph/9908323}}].

\bibitem{Baier:1999db}
R.~Baier and K.~Redlich, {\it Hard thermal-loop resummed pressure of a
  degenerate quark- gluon plasma},  {\em Phys. Rev. Lett.} {\bf 84} (2000)
  2100, [\href{http://xxx.lanl.gov/abs/hep-ph/9908372}{{\tt hep-ph/9908372}}].

\bibitem{Andersen:2002ey}
J.~O. Andersen, E.~Braaten, E.~Petitgirard, and M.~Strickland, {\it {HTL}
  perturbation theory to two loops},  {\em Phys. Rev.} {\bf D66} (2002) 085016,
  [\href{http://arXiv.org/abs/hep-ph/0205085}{{\tt
  hep-ph/0205085}}];
J.~O. Andersen, E.~Petitgirard, and M.~Strickland, {\it Two-loop {HTL}
  thermodynamics with quarks}, 
[\href{http://xxx.lanl.gov/abs/hep-ph/0302069}{{\tt hep-ph/0302069}}].

\bibitem{Blaizot:1999ip}
J.~P. Blaizot, E.~Iancu, and A.~Rebhan, {\it The entropy of the {QCD} plasma},
  {\em Phys. Rev. Lett.} {\bf 83} (1999) 2906,
  [\href{http://xxx.lanl.gov/abs/hep-ph/9906340}{{\tt hep-ph/9906340}}];
{\it Self-consistent hard-thermal-loop
  thermodynamics for the quark-gluon plasma},  {\em Phys. Lett.} {\bf B470}
  (1999) 181, [\href{http://xxx.lanl.gov/abs/hep-ph/9910309}{{\tt
  hep-ph/9910309}}].

\bibitem{Blaizot:2000fc}
J.~P. Blaizot, E.~Iancu, and A.~Rebhan, 
{\it Approximately self-consistent
  resummations for the thermodynamics of the quark-gluon plasma: Entropy and
  density},  {\em Phys. Rev.} {\bf D63} (2001) 065003,
  [\href{http://xxx.lanl.gov/abs/hep-ph/0005003}{{\tt hep-ph/0005003}}].

\bibitem{Blaizot:2001vr}
J.~P. Blaizot, E.~Iancu, and A.~Rebhan, 
{\it Quark number susceptibilities from
  {HTL}-resummed thermodynamics},  {\em Phys. Lett.} {\bf B523} (2001)
  143,
  [\href{http://xxx.lanl.gov/abs/hep-ph/0110369}{{\tt
  hep-ph/0110369}}].

\bibitem{Blaizot:2003tw}
J.-P. Blaizot, E.~Iancu, and A.~Rebhan, 
{\it Thermodynamics of the
  high-temperature quark gluon plasma}, 
[\href{http://xxx.lanl.gov/abs/hep-ph/0303185}{{\tt hep-ph/0303185}}].

\bibitem{Moore:2002md}
G.~D. Moore, {\it Pressure of hot {QCD} at large {$N_f$}},  {\em JHEP} {\bf
  0210} (2002) 055, [\href{http://xxx.lanl.gov/abs/hep-ph/0209190}{{\tt
  hep-ph/0209190}}].

\bibitem{Ipp:2003zr}
A.~Ipp, G.~D. Moore, and A.~Rebhan, {\it Comment on and erratum to `{P}ressure
  of hot {QCD} at large {$N_f$}'},  {\em JHEP} {\bf 0301} (2003) 037,
  [\href{http://xxx.lanl.gov/abs/hep-ph/0301057}{{\tt hep-ph/0301057}}].

\bibitem{Bodeker:1998an}
D.~B{\"o}deker, P.~V. Landshoff, O.~Nachtmann, and A.~Rebhan, {\it
  Renormalisation of the nonperturbative thermal pressure},  {\em Nucl. Phys.}
  {\bf B539} (1999) 233,
  [\href{http://xxx.lanl.gov/abs/hep-ph/9806514}{{\tt hep-ph/9806514}}].

\bibitem{Fodor:2001au}
Z.~Fodor and S.~D. Katz, {\it A new method to study lattice {QCD} at finite
  temperature and chemical potential},  {\em Phys. Lett.} {\bf B534} (2002)
  87, [\href{http://xxx.lanl.gov/abs/hep-lat/0104001}{{\tt
  hep-lat/0104001}}];
{\it Lattice determination of the critical point of
  {QCD} at finite {T} and {$\mu$}},  {\em JHEP} {\bf 0203} (2002) 014,
  [\href{http://xxx.lanl.gov/abs/hep-lat/0106002}{{\tt hep-lat/0106002}}].

\bibitem{Fodor:2002km}
Z.~Fodor, S.~D. Katz, and K.~K. Szabo, {\it 
The {QCD} equation of state at nonzero
  densities: Lattice result}, 
[\href{http://xxx.lanl.gov/abs/hep-lat/0208078}{{\tt hep-lat/0208078}}].

\bibitem{deForcrand:2002ci}
P.~de~Forcrand and O.~Philipsen, {\it The {QCD} phase diagram for small
  densities from imaginary chemical potential},  {\em Nucl. Phys.} {\bf B642}
  (2002) 290, [\href{http://xxx.lanl.gov/abs/hep-lat/0205016}{{\tt
  hep-lat/0205016}}].

\bibitem{Allton:2002zi}
C.~R. Allton {\em et.~al.}, {\it The {QCD} thermal phase transition in the
  presence of a small chemical potential},  {\em Phys. Rev.} {\bf D66} (2002)
  074507,
  [\href{http://arXiv.org/abs/hep-lat/0204010}{{\tt
  hep-lat/0204010}}].

\bibitem{D'Elia:2002gd}
M.~D'Elia and M.-P. Lombardo, {\it Finite density {QCD} via imaginary chemical
  potential},  {\em Phys. Rev.} {\bf D67} (2003) 014505,
  [\href{http://xxx.lanl.gov/abs/hep-lat/0209146}{{\tt hep-lat/0209146}}].

\bibitem{Gavai:2003mf}
R.~V. Gavai and S.~Gupta, {\it
Pressure and non-linear susceptibilities in {QCD}
  at finite chemical potentials}, 
[\href{http://xxx.lanl.gov/abs/hep-lat/0303013}{{\tt hep-lat/0303013}}].

\bibitem{Arnold:1995eb}
P.~Arnold and C.-X. Zhai, {\it The three-loop free energy for high temperature
  {QED} and {QCD} with fermions},  {\em Phys. Rev.} {\bf D51} (1995)
  1906, [\href{http://xxx.lanl.gov/abs/hep-ph/9410360}{{\tt
  hep-ph/9410360}}].

\bibitem{Zhai:1995ac}
C.-X. Zhai and B.~Kastening, {\it The free energy of hot gauge theories with
  fermions through {$g^5$}},  {\em Phys. Rev.} {\bf D52} (1995) 7232,
  [\href{http://xxx.lanl.gov/abs/hep-ph/9507380}{{\tt hep-ph/9507380}}].

\bibitem{Hart:2000ha}
A.~Hart, M.~Laine, and O.~Philipsen, {\it Static correlation lengths in {QCD}
  at high temperatures and finite densities},  {\em Nucl. Phys.} {\bf B586}
  (2000) 443, [\href{http://xxx.lanl.gov/abs/hep-ph/0004060}{{\tt
  hep-ph/0004060}}].

\bibitem{Ginsparg:1980ef}
P.~Ginsparg, {\it First-order and second-order phase transitions in gauge
  theories at finite temperature},  {\em Nucl. Phys.} {\bf B170} (1980) 388;
T.~Appelquist and R.~D. Pisarski, {\it Hot {Y}ang-{M}ills theories and
  three-dimensional {QCD}},  {\em Phys. Rev.} {\bf D23} (1981) 2305;
S.~Nadkarni, {\it Dimensional reduction in hot {QCD}},  {\em Phys. Rev.} {\bf
  D27} (1983) 917;
{\it Dimensional reduction in finite temperature quantum
  chromodynamics. 2},  {\em Phys. Rev.} {\bf D38} (1988) 3287;
N.~P. Landsman, {\it Limitations to dimensional reduction at high temperature},
   {\em Nucl. Phys.} {\bf B322} (1989) 498.

\bibitem{Braaten:1995na}
E.~Braaten, {\it Solution to the perturbative infrared catastrophe of hot gauge
  theories},  {\em Phys. Rev. Lett.} {\bf 74} (1995) 2164,
  [\href{http://arXiv.org/abs/hep-ph/9409434}{{\tt
  hep-ph/9409434}}].

\bibitem{Kajantie:1996dw}
K.~Kajantie, M.~Laine, K.~Rummukainen, and M.~Shaposhnikov, {\it Generic rules
  for high temperature dimensional reduction and their application to the
  standard model},  {\em Nucl. Phys.} {\bf B458} (1996) 90,
  [\href{http://xxx.lanl.gov/abs/hep-ph/9508379}{{\tt hep-ph/9508379}}].

\bibitem{Vuorinen:2002ue}
A.~Vuorinen, {\it Quark number susceptibilities of hot {QCD} up to
  {$g^6\ln(g)$}},  {\em Phys. Rev.} {\bf D67} (2003) 074032,
  [\href{http://xxx.lanl.gov/abs/hep-ph/0212283}{{\tt hep-ph/0212283}}].

\bibitem{Freedman:1977dm}
B.~A. Freedman and L.~D. McLerran, {\it Fermions and gauge vector mesons at
  finite temperature and density. 2. {T}he ground state energy of a
  relativistic electron gas},  {\em Phys. Rev.} {\bf D16} (1977) 1147;
{\it 
3. {T}he ground state energy of a
  relativistic quark gas},  {\em Phys. Rev.} {\bf D16} (1977) 1169.

\bibitem{Baluni:1978ms}
V.~Baluni, {\it Non-abelian gauge theories of {F}ermi systems:
  Quantum-chromodynamic theory of highly condensed matter},  {\em Phys. Rev.}
  {\bf D17} (1978) 2092.

\bibitem{Holstein:1973}
T.~Holstein, R.~E. Norton, and P.~Pincus, {\it {d}e {H}aas-van {A}lphen effect
  and the specific heat of an electron gas},  {\em Phys. Rev.} {\bf B8} (1973)
  2649.

\bibitem{Gan:1993}
J.~Gan and E.~Wong, {\it Non-{F}ermi-liquid behavior in quantum critical
  systems},  {\em Phys. Rev. Lett.} {\bf 71} (1993) 4226.

\bibitem{Chakravarty:1995}
S.~Chakravarty, R.~E. Norton, and O.~F. Sylju{\aa}sen, {\it Transverse gauge
  interactions and the vanquished {F}ermi liquid},  {\em Phys. Rev. Lett.} {\bf
  74} (1995) 1423.

\bibitem{Son:1998uk}
D.~T. Son, {\it Superconductivity by long-range color magnetic interaction in
  high-density quark matter},  {\em Phys. Rev.} {\bf D59} (1999) 094019,
  [\href{http://xxx.lanl.gov/abs/hep-ph/9812287}{{\tt hep-ph/9812287}}].

\bibitem{Brown:2000eh}
W.~E. Brown, J.~T. Liu, and H.-c. Ren, {\it Non-{F}ermi liquid behavior, the
  {BRST} identity in the dense quark-gluon plasma and color superconductivity},
   {\em Phys. Rev.} {\bf D62} (2000) 054013,
  [\href{http://xxx.lanl.gov/abs/hep-ph/0003199}{{\tt hep-ph/0003199}}].

\bibitem{Boyanovsky:2000bc}
D.~Boyanovsky and H.~J. de~Vega, {\it Non-{F}ermi liquid aspects of cold and
  dense {QED} and {QCD}: Equilibrium and non-equilibrium},  {\em Phys. Rev.}
  {\bf D63} (2001) 034016, [\href{http://xxx.lanl.gov/abs/hep-ph/0009172}{{\tt
  hep-ph/0009172}}].

\bibitem{Boyanovsky:2000zj}
D.~Boyanovsky and H.~J. de~Vega, {\it The specific heat of normal, degenerate
  quark matter: Non-{F}ermi liquid corrections},  {\em Phys. Rev.} {\bf D63}
  (2001) 114028, [\href{http://xxx.lanl.gov/abs/hep-ph/0011354}{{\tt
  hep-ph/0011354}}].

\bibitem{Braaten:1996jr}
E.~Braaten and A.~Nieto, {\it Free energy of {QCD} at high temperature},  {\em
  Phys. Rev.} {\bf D53} (1996) 3421,
  [\href{http://xxx.lanl.gov/abs/hep-ph/9510408}{{\tt hep-ph/9510408}}].

\bibitem{Vuorinen:2003fs}
A.~Vuorinen, {\it The pressure of {QCD} at finite temperatures and chemical
  potentials}, 
[\href{http://xxx.lanl.gov/abs/hep-ph/0305183}{{\tt hep-ph/0305183}}].

\bibitem{Kajantie:2002wa}
K.~Kajantie, M.~Laine, K.~Rummukainen, and Y.~Schr{\"o}der, {\it The pressure
  of hot {QCD} up to {$g^6 \ln(1/g)$}},  {\em Phys. Rev.} {\bf D67} (2003)
  105008, [\href{http://xxx.lanl.gov/abs/hep-ph/0211321}{{\tt
  hep-ph/0211321}}].

\bibitem{Blaizot:2003iq}
J.~P. Blaizot, E.~Iancu, and A.~Rebhan, {\it On the apparent convergence of
  perturbative {QCD} at high temperature}, 
[\href{http://xxx.lanl.gov/abs/hep-ph/0303045}{{\tt hep-ph/0303045}}].

\bibitem{Szabo:2003kg}
K.~K. Szabo and A.~I. Toth, {\it Quasiparticle description of the {QCD} plasma,
  comparison with lattice results at finite {T} and {$\mu$},} 
[\href{http://xxx.lanl.gov/abs/hep-ph/0302255}{{\tt hep-ph/0302255}}].

\bibitem{Rebhan:2003wn}
A.~Rebhan and P.~Romatschke, {\it 
{HTL} quasiparticle models of deconfined {QCD}
  at finite chemical potential}, 
[\href{http://xxx.lanl.gov/abs/hep-ph/0304294}{{\tt hep-ph/0304294}}].

\bibitem{Peshier:1999ww}
A.~Peshier, B.~K{\"a}mpfer, and G.~Soff, {\it The equation of state of
  deconfined matter at finite chemical potential in a quasiparticle
  description},  {\em Phys. Rev.} {\bf C61} (2000) 045203,
  [\href{http://xxx.lanl.gov/abs/hep-ph/9911474}{{\tt hep-ph/9911474}}];
{\it From {QCD} lattice calculations
  to the equation of state of quark matter},  {\em Phys. Rev.} {\bf D66} (2002)
  094003, [\href{http://xxx.lanl.gov/abs/hep-ph/0206229}{{\tt
  hep-ph/0206229}}].

\bibitem{Peshier:2002fm}
A.~Peshier, {\it Comment on `{P}ressure of hot {QCD} at large {$N_f$}'},  {\em
  JHEP} {\bf 0301} (2003) 040,
  [\href{http://xxx.lanl.gov/abs/hep-ph/0211088}{{\tt hep-ph/0211088}}].

\bibitem{Fraga:2001id}
E.~S. Fraga, R.~D. Pisarski, and J.~Schaffner-Bielich, {\it Small, dense quark
  stars from perturbative {QCD}},  {\em Phys. Rev.} {\bf D63} (2001) 121702,
  [\href{http://xxx.lanl.gov/abs/hep-ph/0101143}{{\tt hep-ph/0101143}}].

\end{thebibliography}

\providecommand{\href}[2]{#2}\begingroup\raggedright\endgroup

\end{document}